\input amstex

\documentstyle{amsppt}

\magnification=1200

\centerline{\bf On Effective Hamiltonians for Adiabatic Perturbations of}

\centerline{\bf Magnetic Schr\"odinger Operators}

\bigskip

\centerline{Mouez Dimassi, Jean-Claude Guillot and James Ralston}

\bigskip

\noindent{ \bf Abstract.}\it
We construct almost invariant subspaces and the corresponding effective Hamiltonian for magnetic Bloch bands. We also discuss the question of the dynamics related to the effective Hamiltonian.
We assume that the magnetic and electric potentials are slowly varying perturbations of the potential of a
constant magnetic field and a periodic lattice potential, respectively.\rm

\medskip

\centerline{1. \bf Introduction}

In [5] we constructed wave packets for adiabatic perturbations of Schr\"odinger operators in periodic media. The recent work of Panati-Spohn-Teufel, [19], led us to consider the relation of those constructions to effective Hamiltonians. In \S 3 of this article we give a
simple derivation of effective Hamiltonians for these problems.

The main simplification in our method is the omission of the Floquet-Bloch
transformation. This transformation has many nice properties. In particular, it is unitary, and this makes in useful in studying spectral properties of operators. In the work of Helffer-Sj\"ostrand [9]
and G\'erard-Martinez-Sj\"ostrand [7] this transformation was used quite
effectively in the computation of spectra, both of perturbed and effective
Hamiltonians. However, if one is simply interested in effective Hamiltonians,
the Floquet-Bloch transformation requires that one transform the Hamiltonian by a Fourier integral unitary operator only to transform it back at the end of the calculation.
In this article we need to assume that eigenspaces of the unperturbed Hamiltonian depend smoothly on quasi-momentum, and form trivial bundles over a fundamental domain for the dual lattice.

There is also the interesting question of how one interprets the lower order
terms in the effective Hamiltonian. If one considers the propagation of
observables in the Heisenberg picture, it is natural to think of these terms
as lower order corrections to the dynamics. This point of view is adopted in
[19], and it is implicit in [3] and [4]. However, the highest order contributions of these terms to the wave packets are in a precession of the phase. Thus in [5] we did not include them in the dynamics, and did not see how to reconcile our results with those of [3] and [4]. It now appears that the two points of view complement each other instead of conflicting.

\medskip

\centerline{ \bf 2. Preliminaries}

\medskip

The Hamiltonian for an electron in a crystal lattice $\Gamma$ in ${\Bbb R}^3$ in the presence of a
constant magnetic field $\omega=(\omega_1,\omega_2,\omega_3)$ is given by
$$H_0={1\over 2m}\left(-ih{\partial\over\partial x}+e{\omega\times x\over 2}\right)^2 + V(x),\eqno{(1)}$$
where $V$ is a smooth, real-valued potential, periodic with respect to $\Gamma$. Here $m$ and $e$ are the
mass and charge of the electron. To simplify notation we will use units in which $h=2m=e= 1$.

We will assume that $\Gamma$ is generated by the basis $\{e_1,e_2,e_3\}$ for
${\Bbb R}^3$,
$$\Gamma =e_1{\Bbb Z}+e_2{\Bbb Z}+e_3{\Bbb Z},\eqno{(2)}$$
and let $E$ be the fundamental domain $\{\sum_{j=1}^3t_je_j,t_j\in [0,1)\}$.
We will use the dual lattice $\Gamma^*=e_1^*{\Bbb Z}+e_2^*{\Bbb Z}+e_3^*{\Bbb Z}$, where $e_j^*\cdot e_k=2\pi\delta_{jk}$,
with the fundamental domain
$E^*=\{\sum_{j=1}^3t_je_j^*,t_j\in [0,1)\}$.

To realize $H_0$ as a self-adjoint operator in $L^2({\Bbb R}^3)$ we define it
first on the Schwartz functions ${\Cal S}({\Bbb R}^3)$, and then take the
Friedrichs extension. The resulting operator commutes with the magnetic
translations introduced by Zak [24],
$$T_\gamma f(x)=e^{i\langle \omega\times x,\gamma\rangle /2}f(x-\gamma)\eqno{(3)}$$
for $\gamma\in \Gamma$. We assume that
$$\langle \omega, \Gamma\times\Gamma\rangle \subset 4\pi{\Bbb Z}.$$
With this assumption $G=\{T_\gamma, \gamma\in \Gamma\}$ is an abelian
group, and we can reduce $H_0$ by the eigenspaces of $G$, i.e. setting
$${\Cal D}_k=\{u\in H^2_{loc}({\Bbb R^3}), T_\gamma u=e^{-ik\cdot \gamma}u,\gamma\in \Gamma\},\eqno{(4)}$$
considered as a subspace of $L^2(E)$, $H_0$ restricted to ${\Cal D}_k$ is
self-adjoint with compact resolvant. We denote its spectrum by
$$E_1(k)\leq E_2(k)\leq \dots$$
Then by standard results the spectrum of $H_0$ as an operator
in $L^2({\Bbb R}^3)$ is equal to
$$\cup_{k\in E^*}\cup_{m=1}^\infty E_m(k).$$
Note that, since ${\Cal D}_{k+\gamma^*}={\Cal D}_k$ for $\gamma^*\in \Gamma^*$, $E_m(k+\gamma^*)=E_m(k)$.

Standard perturbation theory shows that the function $E_m(k)$ is continuous for $k\in\Bbb R^3$ and
real analytic in a neighborhood of any $k$ such that
$$E_{m-1}(k)<E_m(k)<E_{m+1}(k)\eqno{(5)}$$
The closed interval $\Lambda_m=\cup_{k\in E^*}E_m(k)$ is
known as the \lq\lq m-th magnetic Bloch band" in the spectrum of $H_0$.

In what follows it will be convenient to replace $H_0$ acting on ${\Cal D}_k$ by
$$H_0(k)=e^{-ik\cdot x}H_0e^{ik\cdot x}=\left(-i{\partial\over \partial x}+{\omega\times x\over 2}+k\right)^2 +V(x)$$
with the domain
$$\Cal D=\{u\in H^2_{loc}({\Bbb R}^3), T_\gamma u=u, \gamma\in \Gamma\}.$$
for all $k$. As with $\Cal D_k$, we consider $\Cal D$ as a subspace of $L^2(E)$.\medskip

{\bf Assumption A.} \it For a given $m$ we will assume that $E_m$ satisfies (5) for all $k$.
\noindent \rm
\medskip
Under this assumption we can choose the eigenfunction $\Psi(x,k)$ associated to $E_m(k)$ to be a real-analytic function of $k$ with values in $\Cal D$, such that
 %$\int_E|\Psi(x,k)|^2dx=1$,
$$H_0(k)\Psi(k)=E_m(k)\Psi(k) \hbox{ for all k }\int_E|\Psi(x,k)|^2dx=1.$$
{\bf Assumption B.} We assume that
$$\Psi(x,k+\gamma^*)=e^{i\gamma^*\cdot x}\Psi(x
,k), \gamma^*\in \Gamma^*.$$

This assumption makes the complex line bundle of the eigenspaces a trivial
bundle over the torus
, $\Bbb R^3/\Gamma^*$. In general one has
$$\Psi(x,k+\gamma^*)=e^{i(\gamma^*\cdot x+\theta(k,\gamma^*))}\Psi(x
,k), \gamma^*\in \Gamma^*,$$
where $\theta(k,\gamma^*)$ is real-valued, and determines the structure of the eigenspace bundle.
Since $$\theta(k,m_1e_1^*+m_2e_2^*+m_3e_3^*)=m_1\theta(k,e_1^*)+m_2\theta(k,e^*_2)+
m_3\theta(k,e_3^*)\ \hbox{mod }2\pi,$$
when $\theta(k,\gamma^*)$ is nonzero, the derivatives of $\Psi$ with $k$
will be unbounded and $\Psi$ will not belong to the class of symbols ${\Cal B}$
which we introduce below. Thus we need Assumption B.
\medskip
\medskip
{\bf Remark 1.} \it
The general method of constructing effective Hamiltonians which we give here will apply under the weaker hypothesis:
for a given $m$ there exist $p$ and $q$ such that
$$E_{m-p-1}(k)<E_{m-p}(k)\hbox{ and }E_{m+q}(k)<E_{m+q+1}(k)\hbox{ for all }k.$$
However, in this case the effective Hamiltonian will be a matrix operator acting on functions with values in $\Bbb C^{p+q+1}$, as in [6],[7], [9]  and [19]. \rm
\medskip
\centerline{\bf 3. Main Result}
\medskip
The adiabatically perturbed Hamiltonian is
$$H_\epsilon =\left(-i{\partial\over \partial x}+{\omega\times x\over 2}+A(\epsilon x)\right)^2 +V(x)+W(\epsilon x),$$
where $W$ and $A=(A_1,A_2,A_3)$ are smooth, and bounded together with all of their derivatives.
As before, we define $H_\epsilon$ first on ${\Cal S}(\Bbb R^3)$, and then take the Friedrichs
extension to get a self-adjoint operator in $L^2(\Bbb R^3)$.

The essential step in applying multi-scale techniques is
simply to consider $y=\epsilon x$ as a new independent variable in $H_\epsilon$. Let
$$\tilde H_\epsilon =\left(-i{\partial\over \partial x}-i\epsilon{\partial\over\partial y}+{\omega\times x\over 2}+A(y)\right)^2 +V(x)+W(y).$$
Then, for $u(x,y)$ we can define $w(x)=u(x,\epsilon x)$ and conclude that
$$[\tilde H_\epsilon u](x,\epsilon x)=[H_\epsilon w](x).\eqno{(6)}$$
The identity (6) enables us to solve the Schr\"odinger equation for $H_\epsilon$ uniformly in $\epsilon$ by
solving the Schr\"odinger equation for $\tilde H_\epsilon$ uniformly in $(y,\epsilon)$. The latter might sound
more difficult, but it turns out not to be.

Let $\Cal B$ denote the subspace of $C^\infty(\Bbb R^3\times\Bbb R^3\times \Bbb R^3)$ consisting of
functions of the form
$$P(x,y,k,\epsilon)=P_0(x,y,k)+\epsilon P_1(x,y,k)+\cdots+\epsilon^NP_N(x,y,k)$$
such that $P(x+\gamma,y,k,\epsilon)=e^{i\langle \omega\times x,\gamma\rangle /2}P(x,y,k,\epsilon)$ and
$$\sup_{y,k}\Vert \partial_y^\alpha\partial_k^\beta P_j(\cdot,y,k)\Vert_{L^2(E)}<\infty, \hbox { for all } \alpha, \beta\in {\Bbb N}^3.$$
To $P\in \Cal B$ we associate the $\epsilon$-pseudo-differential operator
$$P(x,y,\epsilon D_y,\epsilon)f(x,y,\epsilon)=(2\pi\epsilon)^{-3}\int e^{ik\cdot(y-z)/\epsilon}P(x,y,k,\epsilon) f(z)dzdk, f\in \Cal S(\Bbb R^3).$$
Note that here we are using the standard quantization -- as opposed to the Weyl quantization.
Our main result is the following:
\medskip
\noindent {\bf Theorem.} \it
For every $N\in \Bbb N$ there exist $P_N=F_0 + \epsilon F_1+\cdots +\epsilon^NF_N\in \Cal B$ and
$H_{eff}^N=h_0+\epsilon h_1+\cdots +\epsilon^Nh_N\in \Cal B$ (independent on $x$) such that
$$\tilde H_\epsilon(P_N(x,y,\epsilon D_y,\epsilon) u)-P_N(x,y,\epsilon D_y,\epsilon)H^N_{eff}(y,\epsilon D_y)u =\Cal O(\epsilon^{N+1})\eqno{(7)}$$
for $u\in \Cal S(\Bbb R^3)$. Moreover, considered as an operator from $L^2(\Bbb R^3)$ into $L^2(E\times \Bbb R^3)$, $P_N$
is approximately isometric, i.e. $P^*_NP_N=I+\Cal O(\epsilon^{N+1})$.
\rm
\medskip
\noindent We interpret $H_{eff}^N$ as the effective Hamiltonian up to order $\epsilon^N$. The leading term in its symbol
is $h_0(y,k)=E_m(k+A(y))+W(y)$. This is the well-known \lq\lq Peierls substitution", [20].
The
symbol of $h_1$ is also quite interesting, and we discuss it in \S 4.
\medskip
\noindent {\bf Proof.} As in \S 2 it will be convenient to work with
$$\tilde H_\epsilon(k)=e^{-ik\cdot x}\tilde H_\epsilon e^{ik\cdot x}=\left(-i{\partial\over \partial x}-i\epsilon{\partial\over\partial y}+{\omega\times x\over 2}+A(y)+k\right)^2 +V(x)+W(y),$$
acting on functions in $\Cal B$, in place of $\tilde H_\epsilon$. Note that
$$\tilde H_\epsilon(P_N(x,y,\epsilon D_y,\epsilon)u)=(2\pi \epsilon)^{-3}\int e^{ik\cdot(y-z)/\epsilon}\tilde H_\epsilon(k)P_N(x,y,k,\epsilon)u(z)dzdk.$$

The Hamiltonian $\tilde H_\epsilon(k)$ can be written as $\tilde H_\epsilon(k) =\tilde H_0(k)+\epsilon \tilde H_1(k) +
\epsilon^2 \tilde H_2(k)$, where
$$\tilde H_0(k)=(-i\partial_x +{\omega\times x\over 2}+k + A(y))^2 +V(x)+W(y)=H_0(k+A(y))+W(y)$$
$$\tilde H_1(k)=-2i(-i\partial_x +{\omega\times x\over 2}+k +A(y))\cdot\partial_y -i\partial_y\cdot A(y),\eqno{(8)}$$
and $\tilde H_2(k)=-\Delta_y.$

We will simply construct the symbols of the pairs $(h_0,F_0)$, $(h_1, F_1), \dots$, successively so that (7) holds to order $\Cal O(\epsilon^{N+1})$ in the $\epsilon$-pseudo-idfferential calculus.
To cancel the order zero terms in (7) we set $h_0(y,k)=E_m(k+A(y))+W(y)$ and $F_0(x,y,k)=\Psi(x,k+A(y))$.
Then, since the symbol of $F_0(x,y,\epsilon D_y)h_0(y,\epsilon D_y)$ is
$$F_0(x,y,k)h_0(y,k)+{\epsilon\over i}{\partial F_0\over\partial k}(x,y,k)\cdot {\partial h_0\over\partial y}(y,k)+\Cal O(\epsilon^2),$$
we must have
$$(\tilde H_0(k)-h_0)F_1-{1\over i}{\partial F_0\over\partial k}\cdot {\partial h_0\over\partial y}+(\tilde H_1(k)-h_1)F_0=0.\eqno{(9)}$$
By the
Fredholm Alternative in $ L^2(E)$, we can solve (9) for $F_1$ if and only if$$\langle F_0(\cdot,y,k),-{1\over i}{\partial F_0\over\partial k}(\cdot,y,k)\cdot {\partial h_0\over\partial y}+(\tilde H_1(k)-h_1(y,k))F_0(\cdot, y,k)\rangle=0,$$
where $\langle\cdot,\cdot\rangle$ denotes the inner product in $L^2(E)$.
Hence, we choose
$$h_1(y,k)= \langle F_0(\cdot,y,k),{-1\over i}{\partial F_0\over\partial k}(\cdot,y,k)\cdot {\partial h_0\over\partial y}(y,k)+\tilde H_1(k)F_0(\cdot, y,k)\rangle\hbox{ and }\eqno{(10)}$$
$$ F_1(x,y,k)=(\tilde H_0(k)-h_0(y,k))^{-1}({1 \over i}{\partial F_0\over\partial k}\cdot {\partial h_0\over\partial y}+h_1F_0-\tilde H_1(k)F_0) + a_1(y,k)F_0,\eqno{(11)}$$
where $(\tilde H_0(k)-h_0(y,k))^{-1}$ denotes the inverse which maps the orthogonal complement of $F_0(\cdot,y,k)$ in
$L^2(E)$ onto itself. We recall that
$$
\tilde H_0(k)-h_0(y,k)=H_0(k+A(y))-E_m(k+A(y)).$$
We determine $a_1(y,k)$ by the requirement that $P^*_NP_N=I+\Cal O(\epsilon^{N+1})$. Given
$f,g\in \Cal S(\Bbb R^3)$, this implies
$$\int_{E}dx\int_{\Bbb R^3}[\overline{(F_0(x,y,\epsilon D_y)+\epsilon F_1(x,y,\epsilon D_y))f}][(F_0(x,y,\epsilon D_y)+\epsilon F_1(x,y,\epsilon D_y))g]dy=$$
$$
\int_{\Bbb R^3}\overline f gdy +\Cal O(\epsilon^2).\eqno{(12)}$$
Since $\int_{E}|F_0(x,y,k)|^2dx=1$ for all $(k,y)$, one can calculate Re$\{ a_1(y,k)\}$ from (12), and see that it
is smooth and bounded, by the pseudo-differential calculus. We choose Im$\{a_1(y,k)\}=0$.

The calculation of $(F_j,h_j)$ for $j>1$ proceeds in the same manner: we calculate the terms of order $\epsilon^j$ in the symbol of the left hand side of (7) which come from $(F_l,h_l)$ for $l<j$ and choose $(F_j,h_j)$ so that  the left hand side of (7) has the desired form up to terms of order $\epsilon^{j+1}$. At each stage we use the Fredholm Alternative in
$L^2(E)$, and $F_j$ is only determined modulo a term of the form $a_j(y,k)F_0(x,y,k)$. Then we choose the real part of $a_j(y,k)$ so that $P^*_jP_j=I+\Cal O(\epsilon^{j+1})$, and take the imaginary part of $a_j$ to be
zero. Continuing in this way we complete the proof of the Theorem.
\medskip
{\bf Remark 2}
\it If we set $\Pi_N=P_NP_N^*$, then $\Pi_N$ is approximately a projection:
$\Pi_N^2=\Pi_N+\Cal O(\epsilon^{N+1})$ and $\Pi_N=\Pi_N^*$.
The equation (7) implies that
$$P_N ^*\tilde H_\epsilon=H_{eff}P_N^* +\Cal O(\epsilon^{N+1}).$$
Hence
$$P_NP_N^*\tilde H_\epsilon=P_N H_{eff} P_N^*+\Cal O(\epsilon^{N+1}).$$
If we replace $P_N H_{eff}$ in the equality above by the left hand side of (7),we obtain:
$$P_NP_N^* \tilde H_\epsilon=\tilde H_\epsilon P_NP_N^*+ \Cal
O(\epsilon^{N+1}).$$
Thus $\Pi_N$ is a
projection which commutes  with
$\tilde H_\epsilon$ to order $\Cal O(\epsilon^{N+1})$ as in [19].

In [19], the construction of the almost invariant subspaces is based on the method of  Nenciu-Sordoni [17] and Sordoni[18] (see also [10], [14],[15]). This method is heavily
related  to the construction of Moyal projections.

\rm

\medskip
\centerline{\bf 4. Relations with Previous Work}
\medskip
To relate the results here to what has already been done we need to complete the calculation of the effective Hamiltonian up to terms of order $\epsilon^2$, i.e. to compute the symbol $h_1(y,k)$ from (10). The explicit computation of
$$\langle F_0(\cdot,y,k),\tilde H_1(k)F_0(\cdot, y,k)\rangle=_{def} E_1(y,k)$$
is contained in the computations in [5] (it is also in
[8] with a small error -- see Remark 1 in [5]). When one replaces $k(y,s)$ and $\Psi(x,y,s)$ in [5, pp. 7601-3] by $k+A(y)$ and
$\Psi(x,k+A(y))$ respectively, $E_1(y,k)$ is \lq\lq$ih$" in the notation of [5, (25)] and we have
$$E_1={1\over 2i}{\partial \over \partial y}\cdot {\partial E_m(\tilde k)\over\partial k}-L\cdot B -i\langle \Psi(\cdot,\tilde k),{\partial\Psi(\cdot,
\tilde k)\over \partial y}\rangle\cdot
{\partial E_m(\tilde k)\over\partial k},$$
where $\tilde k=k+A(y)$.
Here $B(y)=\nabla\times A(y)$ and $L=$
$$L=\hbox{Im}\left(\langle M(y,k)
{\partial \Psi\over \partial k_2},{\partial \Psi\over\partial k_3}\rangle,
\langle M(y,k) {\partial \Psi\over \partial k_3},{\partial \Psi\over\partial k_1}\rangle_,\langle M(y,k){\partial \Psi\over \partial k_1},{\partial \Psi\over\partial k_2}\rangle\right).$$
with $M(y,k)=\tilde H_0(k)-h_0(y,k)$.
The vector $L$ is an angular momentum and $L\cdot B$ contributes the \lq\lq Rammal-Wilkinson" term to the energy, cf. [1].
Adding the additional term from (10) to $E_1$ to obtain $h_1$, we obtain
$$h_1(y,k)={1\over 2i}{\partial \over \partial y}\cdot {\partial E_m(k+A(y))\over\partial k}-L\cdot B -i\langle \Psi(\cdot,k+A(y),\dot\Psi(\cdot,k+A(y))\rangle,\eqno{(13)}$$
where
$$\dot\Psi(x,k+A(y))={\partial\Psi(x,k+A(y))\over \partial y}\cdot \dot y+{\partial\Psi(x,k+A(y))\over \partial k}\cdot \dot k$$
and $\dot y$ and $\dot k$ are defined by the Hamiltonian system
$$\dot y={\partial (E_m(k+A(y))+W(y))\over \partial k}\qquad\qquad \dot k =-{\partial (E_m(k+A(y))+W(y))\over \partial y}.$$
Thus one recognizes $i\langle \Psi(\cdot,k+A(y)),\dot\Psi(\cdot,k+A(y))\rangle$ as the term generating the Berry phase precession,
cf. [13], [22]. Comparing (13) with [19, (22)] (in the case $l=1$), one sees that they agree completely when one takes into account the difference in the
choice of sign in the magnetic potential, $A(y)$, and the use of Weyl quantization in [19]. The sign of the Berry phase term in (13) may appear inconsistent
 with [5, (29)], but it is not. In [5] $(\dot y,\dot k)$ was the vector field from the Hamiltonian $-E_m(k+A(y))$.

In [5] and [8] instead of introducing effective Hamiltonians we constructed wave packets. These packets are nonetheless related to effective Hamiltonians
in that one can compute what the effective Hamiltonian must be -- assuming that there is one -- from the packets. To see this one can
proceed as follows. The packets have the form (here $s=\epsilon t$ and $W(y)=0$)
$$u(x,y,s,\epsilon)=e^{i\phi(y,s)/\epsilon}[f(y,s)\Psi(x,{\partial\phi\over\partial y}+A(y))+\Cal O(\epsilon)],$$
where $\phi$ and $f$ are solutions of
$${\partial \phi\over \partial s}=E_m({\partial\phi\over\partial y}+A(y))\hbox{ and }{\partial f\over\partial s}={\partial E_m \over\partial k}({\partial \phi\over\partial y}+A(y))\cdot {\partial f\over \partial y}+(D-iL\cdot B+\langle \Psi,\dot\Psi\rangle)f.\eqno{(14)}$$
Here all functions of $(k,y)$ are evaluated at $k=\tilde k(y,s)=\partial_y\phi$, and $D=(1/2)\partial_y\cdot (\partial_kE_m(\partial_y\phi+A(y)))$. Assuming that the evolution of $f$ is governed by an effective Hamiltonian
$H_{eff}=h_0(y,\epsilon D_y)+\epsilon h_1(y,\epsilon D_y)+\Cal O(\epsilon^2)$, we must have (on bounded intervals in $s$)
$$[e^{isH_{eff}/\epsilon}e^{i\phi(\cdot,0)/\epsilon}f(\cdot,0)](y,s)=e^{i\phi(y,s)/\epsilon}f(y,s)+\Cal O(\epsilon^2).\eqno{(15)}$$
Differentiating (15) with respect to $s$, one concludes
$${i\over \epsilon}H_{eff}(e^{i\phi(y,s)/\epsilon}f(y,s))=({i\over\epsilon}{\partial\phi\over \partial s}f+{\partial f\over \partial s})e^{i\phi(y,s)/\epsilon} +\Cal O(\epsilon).\eqno{(16)}$$
Using the symbol expansion from the pseudo-differential calculus
$$e^{-i\phi/\epsilon}H_{eff}(e^{i\phi/\epsilon}f)=h_0(y,\tilde k)+$$
$$\epsilon[{1\over i}{\partial h_0\over\partial k}(y,\tilde k)\cdot {\partial f\over\partial y}(y)+{1\over 2i}\sum_{j,l} {\partial^2h_0\over\partial k_j\partial k_l}(y,\tilde k){\partial^2\phi\over\partial y_j\partial y_l}(y)+h_1(y,\tilde k)]f(y) +\Cal O(\epsilon^2).\eqno{(17)}$$
Substituting (17) into (16) and comparing the result with (14), one recovers the formulas given earlier for $h_0(y,k)$ and $h_1(y,k)$.

In [8] we were unable to reconcile our results with those of Chang and Niu, see [3], [4] and also [23]). We thought that this might have resulted from
different choices of scales. This is partially true, since Chang and Niu do not distinguish the scale $y=\epsilon x$, but, as Panati, Spohn and
Teufel point out in [19], the differences largely disappear when one considers the Heisenberg formulation of quantum dynamics. Letting $a(y,\epsilon D_y)$ be an observable, the propagation of $a$ in the Heisenberg picture is given by
$$a(s)=e^{-isH_{eff}/\epsilon}ae^{isH_{eff}/\epsilon}\hbox{ or }\epsilon{da\over ds}=i[a,H_{eff}].$$
If one considers this propagation at the symbol level, then the symbol of $a$ is propagating along the trajectories of the Hamiltonian system
$$\dot y={\partial H_{eff}\over\partial k}\qquad\qquad \dot k=-{\partial H_{eff}\over \partial y}.$$
Hence one can consider the contribution of $h_1$ as an order $\epsilon$ correction to the classical Peierls dynamics arising from $h_0$. This is
the point of view taken in [19]. However, it is worth noting that the wave packets are propagating along the trajectories from $h_0$ with a precession in
their phases arising from (the imaginary part of) $h_1$.
\medskip\noindent
{\bf Acknowledgements}
We wish to thank Professors
Panati, Spohn and Teufel for sending us the preprint  version of [19] and
for several helpful discussions of this work.
 We especially thank Stefan Teufel for pointing out
the necessity of Assumption B here.
%We would like to thank A. Martinez for pointing out many references for us.
\medskip
\centerline{\bf References}
\noindent [1] J. Bellissard and R. Rammal, {\it An algebric semi-classical
approach to Bloch electrons in a magnetic field }
J. Physique France {\bf 51}(1990), 1803.
\medskip
\noindent [2] V. S. Buslaev, {\it Semi-classical
approximation for equations with periodic
coefficients.}
Russian. Math. Surveys, {\bf 42} (1987), 97--125.
\medskip
\noindent [3] M. C. Chang and Q. Niu, {\it Berry phase, hyperorbits, and
the Hofstadter spectrum.}
Phys. Rev. lett. {\bf 75}(1996), 1348-1351.
\medskip
\noindent [4] M. C. Chang and Q. Niu, {\it Berry phase, hyperorbits,
and the Hofstadter spectrum: semiclassical in magnetic Bloch bands}
Phys. Rev. B {\bf 53}(1996) 7010-7022.
\medskip
\noindent [5] M. Dimassi, J.-C. Guillot and J. Ralston,
{\it
Semi-Classical Asymptotics in Magnetic Bloch Bands.
}
J. Phys. A: Math. G., {\bf 35} (2002), 7597--7605 .
\medskip
\noindent [6] M. Dimassi and J. Sj\"ostrand, {\it Spectral asymptotics in the
semi-classical limit.
}
London Math. Soc. Lecture Note Series, 268.
Cambridge University Press, Cambridge, 1999.
\medskip
\noindent [7] C. G\'erard,
A. Martinez and J. Sj\"ostrand, {\it A Mathematical
Approach to the effective Hamiltonian in
perturbed periodic Problems. }
Commun. Math.
Phys., {\bf 142} (1991), 217--244.
\medskip
\noindent [8] J.-C. Guillot,
J. Ralston and E. Trubowitz, {\it Semi-classical
methods in solid state physics. }
Commun.
Math. Phys., {\bf 116} (1988), 401--415.
\medskip
\noindent [9] B. Helffer and J. Sj\"ostrand, {\it On diamagnetism and the de Haas - van Alphen effect.} Annales I.H.P.
(Physique th\'eorique) {\bf 52} (1990), 303-375.
\medskip
\noindent [10] B. Helffer and J. Sj\"ostrand,
{\it Analyse semiclassique pour l'\'equation de Harper II.}
M\'em. S.M.F. {\bf 40}, 139p, (1990).
\medskip
\noindent [11] F. H\^overmann, H. Spohn and S. Teufel, {\it
Semi-classical limit for the Schr\"odinger equation with a short scale periodic potential. }
Comm. Math. Phys. {\bf 215}
(2001), no. 3, 609--629.
\medskip
\noindent [12] W. Horn, {\it Semi-classical construction in solid state physics.}
Commun. P.D.E. {\bf 16}(1993) 255-290.
\medskip
\noindent [13] M. Kohmoto, {\it Berry's phase of Bloch electrons in electromagnetic fields.}
J. Phys. Soc. Japan {\bf 62}(1993), 659-663.
\medskip
\medskip
\noindent [14] A. Martinez and  V. Sordini
{\it A general reduction scheme for the time-dependent Born-Oppenheimer approximation. }  C. R.
   Math. Acad. Sci. Paris {\bf 334}  (2002), no. 3, 185--188

\medskip
\noindent [15] A. Martinez and  V. Sordoni,
On the Time-Dependent Born-Oppenheimer Approximation with Smooth Potential.
mp-arc 01-37.

\medskip
\noindent [16]  G. Nenciu, {\it
Bloch electrons in a magnetic field: rigorous justification of the Peierls-Onsager effective Hamiltonian.}
 Lett. Math.
   Phys. {\bf 17} (1989).

\medskip
\noindent [17]  G. Nenciu and    V. Sordoni,
       {\it Semiclassical limit for multistate Klein-Gordon systems: almost invariant subspaces and scattering theory.}  mp-arc 01-36.

\medskip
\noindent [18] V. Sordoni
       {\it Reduction Scheme for Semiclassical Operator-valued Schr\"odinger Type Equation and Application to Scattering,}
Commun in Paritial Diff. Equ. {\bf 28} 1221-1236 (2003).

\medskip
\noindent [19] G. Panati, H. Spohn and  S. Teufel,
{\it Effective dynamics for Bloch electrons: Peierls substitution and beyond.}
mp-arc 02-516 (to appear in Commun. Math. Phys.).
\medskip
\noindent [20] R. Peierls, {\it Zur Theorie des diamagnetimus von leitungselektronen } Z. Phys. {\bf 80}(1933), 763-791.
\medskip
\noindent [21] J. Ralston, {\it Magnetic breakdown} Ast\'erisque {\bf 210}(1992), 263-2282.
\medskip
\noindent [22] B. Simon, {\it Holonomy, the quantum adiabatic theorem, and Berry's phase}, Phys. Rev. Lett. {\bf 51}(1983), 2167-2170.
\medskip
\noindent [23] G. Sundaram and Q. Niu, {\it Wave packet dynamics in slowly perturbed crystals: Gradient corrections and Berry phase effects,}
Phys. Rev. B {\bf 59}(1999), 14915-14925.
\medskip
\noindent [24] J. Zak, {\it Dynamics of electrons in solids in external fields.} Phy. Rev. {\bf 168}(1968), 686-695.
\medskip
M. Dimassi and J.-C. Guillot

D\'epartement de Math\'ematiques, Universit\'e Paris 13, Villetaneuse, France

email: dimassi\@math.univ-paris13.fr

email: guillot\@math.univ-paris13.fr
\bigskip
J. Ralston

University of California, Los Angeles, CA 90095, USA

email: ralston\@math.ucla.edu
\end